\documentclass[fleqn,usenatbib]{mnras}

\usepackage{newtxtext,newtxmath}

\usepackage[T1]{fontenc}

\DeclareRobustCommand{\VAN}[3]{#2}
\let\VANthebibliography\thebibliography
\def\thebibliography{\DeclareRobustCommand{\VAN}[3]{##3}\VANthebibliography}

\usepackage{graphicx}	
\usepackage{amsmath}	

\title[Centaurus~A Inner Lobes]{Centaurus~A Inner Lobes – I. Hydrodynamic modeling of a Precessing Jet}

\author[Rita C. Anjos and Brian Reville]{
Rita C. Anjos,$^{1,2}$\thanks{E-mail: ritacassia@ufpr.br}
Brian Reville,$^{1}$
\\
$^{1}$Max-Planck-Institut für Kernphysik, Saupfercheckweg 1, D-69117 Heidelberg, Germany\\
$^{2}$Departamento de Engenharias e Exatas, Universidade Federal do Paraná (UFPR), Pioneiro, 2153, 85950-000 Palotina, PR, Brazil
}

\date{Accepted XXX. Received YYY; in original form ZZZ}

\pubyear{2015}

\begin{document}
\label{firstpage}
\pagerange{\pageref{firstpage}--\pageref{lastpage}}
\maketitle

\begin{abstract}
We present a numerical investigation into the precessing jets of the inner lobes of Centaurus~A, focusing on their dynamical evolution and interaction with the surrounding medium. Using three-dimensional relativistic hydrodynamic simulations, we model the development of large-scale jet and lobe structures driven by precession. Our setup incorporates physically motivated parameters to reproduce observed morphological features. We compare the resulting structures from our simulations with observed radio images of Centaurus~A, particularly focusing on the S-shaped morphology, the distribution of bright emission regions, and the observed asymmetry between the northern and southern lobes. Our findings indicate a precession period of 1.8 Myr, which reproduces observational characteristics. This study explores the role of jet precession in shaping the inner lobes of Centaurus~A.
\end{abstract}




\section{Introduction}

Centaurus~A (NGC~5128) is the closest radio-loud active galaxy to the Milky Way \citep{1998A&ARv...8..237I}, located at a distance of $3.8 \pm 0.1\,\mathrm{Mpc}$ \citep{2010PASA...27..457H}. It serves as a uniquely accessible laboratory for studying jet–intracluster medium (ICM) interactions, feedback processes, and the evolution of radio lobes across multiple spatial scales. Its multi-phase ICM, disturbed stellar shell structures \citep{1983ApJ...272L...5M}, and complex radio morphology \citep{1992ApJ...395..444C, 1983ApJ...273..128B} have made it an archetype of low-power Fanaroff–Riley type I (FRI) radio galaxies.

Centaurus~A exhibits an intricate radio structure spanning nearly six orders of magnitude in scale, from a sub-parsec Very Long Baseline Interferometry (VLBI) jet \citep{1983ApJ...266L..93P} to $\sim 500\,\mathrm{kpc}$ outer lobes \citep{1993A&A...269...29J}. Originating from the core are the inner lobes, which together extend approximately $11\,\mathrm{kpc}$ and are aligned at a position angle of $55^{\circ}$ counterclockwise from the north–south axis \citep{2015MNRAS.447.1001W, 2013MNRAS.436.1286M, 1981ApJ...251..523S, 1992ApJ...395..444C}. The inner kiloparsec jet in Centaurus~A is highly collimated, with an estimated kinetic power of $\sim 1 \times 10^{43}\,\mathrm{erg\,s^{-1}}$, and exhibits signs of deceleration along its path, consistent with entrainment models \citep{2015MNRAS.447.1001W}. Its physical age is estimated at approximately $2\,\mathrm{Myr}$, based on dynamical modeling and spectral aging analyses \citep{2009MNRAS.395.1999C, 2013A&A...558A..19W, 2015MNRAS.447.1001W}, and references therein. Observations at radio and X-ray wavelengths reveal a rich substructure within the inner jet, including multiple compact knots on kiloparsec scales~\citep{2002ApJ...569...54K, 2003ApJ...593..169H, 2006ApJ...641..158K, 2008ApJ...673L.135W, 2009AJ....138..808T, 2010ApJ...708..675G, 2015MNRAS.447.1001W, 2025MNRAS.539.3697D}. Radio knots with higher apparent proper motions exhibit relatively weak X-ray emission, suggesting variations in particle acceleration or emission mechanisms along the jet~\citep{2010ApJ...708..675G, 2015MNRAS.447.1001W}. Furthermore, H.\,E.\,S.\,S. observations of the kiloparsec-scale jet in Centaurus A resolve extended TeV emission precisely aligned with its X-ray jet, confirming that the same population of ultrarelativistic electrons produces both the synchrotron X-rays and the diffuse very-high-energy (VHE) gamma rays \citep{2020Natur.582..356H}. Beyond the northern inner lobe lies the northern middle lobe (NML), the brightest section of the northern structure, located about $30\,\mathrm{kpc}$ from the core and oriented at a position angle of $45^{\circ}$ \citep{1999MNRAS.307..750M}. A direct southern counterpart to the NML has not been detected in total intensity images \citep{2011ApJ...740...17F}, although \citet{2013ApJ...764..162O} identified a region of high fractional polarization in the southern lobe that may represent its analog \citep{2013MNRAS.436.1286M}. The outer lobes are the largest structures, spanning a total extent of $\sim$ $500\,\mathrm{kpc}$. The northern outer lobe initially extends at a position angle between $0^{\circ}$ and $20^{\circ}$~\citep{1965AuJPh..18..589C, 1983PASA....5..241H, 2011ApJ...740...17F}. The southern outer lobe also changes orientation with distance, though more gradually \citep{2013MNRAS.436.1286M}.

\begin{figure*}
   \centering
   \includegraphics[width=0.6\textwidth]{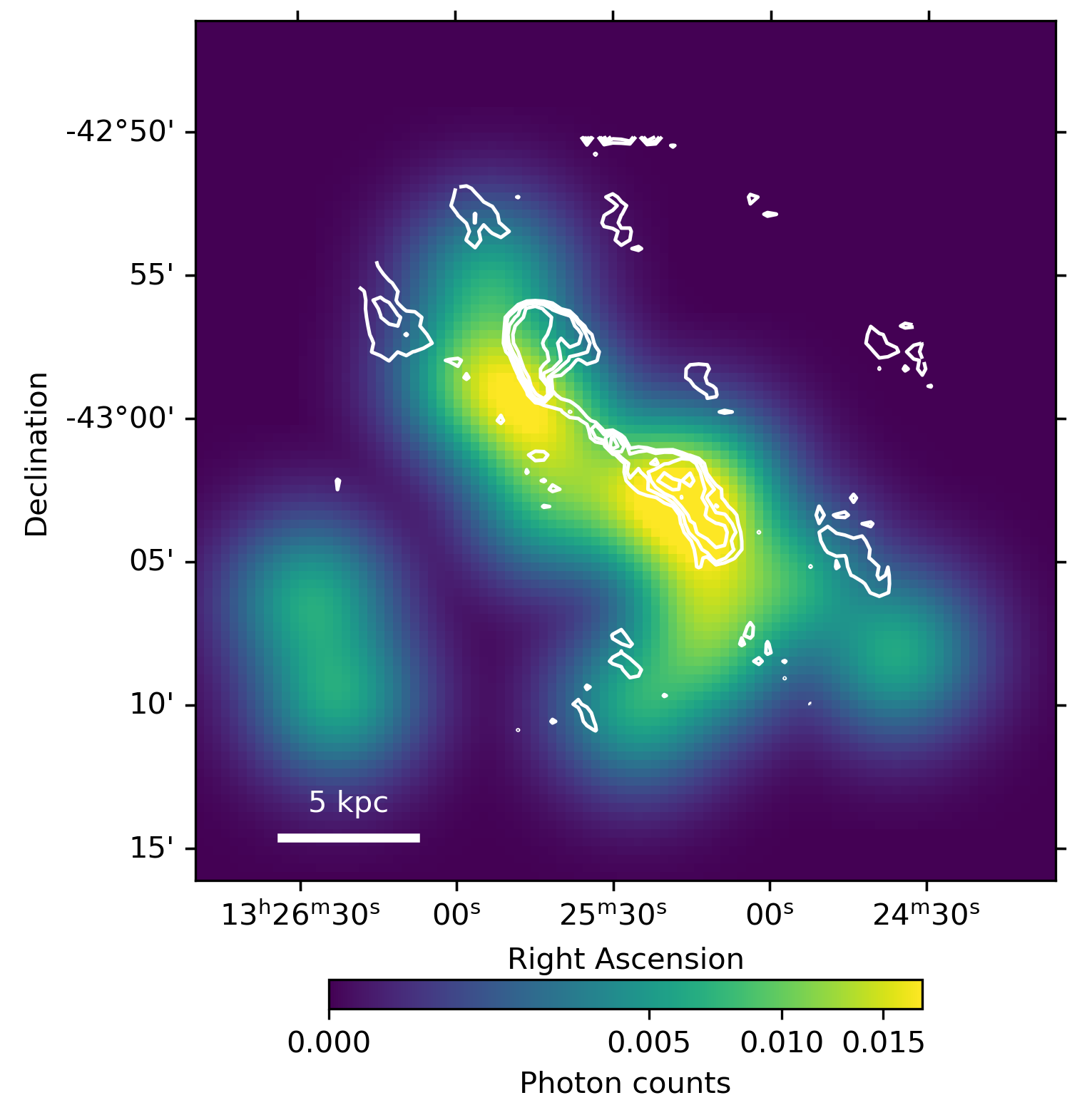}
 \caption{%
    Smoothed {\it Fermi}-LAT count image of PSF3 CLEAN events in the $6.5 - 300\,\mathrm{GeV}$ band over a $3^\circ$ region centered on the Centaurus~A core, shown in equatorial (RA/Dec) coordinates \citep{2019MNRAS.483.3444P}. The image has been convolved with a $0.06^\circ$ Gaussian kernel and is overlaid with white contours of $1.4\,\mathrm{GHz}$ VLA radio continuum emission tracing the north-eastern and south-western inner lobes \citep{1992ApJ...395..444C}. The double-peaked gamma-ray morphology closely follows these radio lobes, indicating spatially resolved high-energy emission.}
  \label{fig:cenA_inner_lobes}
\end{figure*}

High directional \textit{Fermi}-LAT observations have spatially resolved GeV gamma-ray emission from the $1.4\,\mathrm{GHz}$ inner lobes of Centaurus~A~\citep{2019MNRAS.483.3444P} (Figure~\ref{fig:cenA_inner_lobes}). The double-peaked morphology in the $6.5 - 300\,\mathrm{GeV}$ band closely follows the Very Large Array (VLA) radio contours of the north-eastern and south-western lobes, strongly suggesting a common population of relativistic electrons. In a leptonic scenario, these electrons are accelerated at internal shocks and jet–ICM interaction via synchrotron emission from radio to X-ray frequencies and upscatter cosmic microwave background (CMB) and infrared photons to GeV–TeV energies through inverse Compton (IC) processes. The combination of a very weak magnetic field (of order microgauss) and highly relativistic electrons indicates that the plasma is particle-dominated, emphasizing the significance of jet–lobe interactions in influencing the morphology of inner lobes and broadband emission~\citep{2016A&A...595A..29S}. Moreover, by integrating the gamma-ray flux above $6.5\,\mathrm{GeV}$, \citet{2019MNRAS.483.3444P} reported $F_{\gamma,\,\rm SW}=(11.2\pm3.8)\times10^{-11}\,$ph\,cm$^{-2}$\,s$^{-1}$ for the south-western lobe versus $F_{\gamma,\,\rm NE}=(5.2\pm2.7)\times10^{-11}\,$ph\,cm$^{-2}$\,s$^{-1}$ for the north-eastern lobe, revealing a significant gamma-ray asymmetry that complements the radio-morphological contrast~\citep{2019MNRAS.483.3444P}.  

While shock-driven and IC processes account well for the GeV emission, they alone struggle to sustain the ultrarelativistic electron energies implied by the extended X-ray and TeV components. Recent theoretical work therefore invokes distributed shear acceleration as a compelling mechanism in Centaurus~A’s inner lobes. In this model, velocity-shear stratification naturally arises in the collimated flow of relativistic jets~\citep{2019ApJ...886L..26R,2021MNRAS.505.1334W,2022ApJ...933..149R}, and electrons gain energy through scattering off magnetic inhomogeneities embedded in differentially sheared layers, yielding a power-law spectrum with an exponential-like cutoff. For Centaurus~A, Fokker–Planck modeling predicts a steep electron spectrum with a cutoff at tens of TeV, consistent with the observed synchrotron X-rays and TeV inverse-Compton emission; the implied magnetic field of order $B\sim17\,\mu\mathrm{G}$ aligns with radio/X-ray morphology~\citep{2021MNRAS.505.1334W}.

The gamma-ray and radio asymmetry in the inner lobes of Centaurus~A, especially the enhanced northern radio emission, has been the subject of numerous studies and remains poorly understood. Combined observations from the Australia Telescope Compact Array (ATCA) and the VLA reveal pronounced differences in brightness and morphology between the inner lobes, along with bending and misalignment of the jet axis relative to both the small-scale jet and the outer lobes \citep{1983ApJ...273..128B, 1992ApJ...395..444C, 1999MNRAS.307..750M, 2002ApJ...569...54K, 2003ApJ...593..169H, 2003ApJ...592..129K}. Several mechanisms have been proposed to explain these features. \citet{1983ApJ...273..128B} reported that the northern inner lobe (NIL) is significantly broader than the southern one, despite similar jet widths, suggesting that the asymmetry arises within the lobes due to environmental influences. Supporting this interpretation, \citep{2003ApJ...592..129K} identified a bright X-ray shell surrounding only the southwest lobe, likely caused by supersonic expansion into a denser medium. Both studies suggest that small-scale variations in the ICM can significantly affect jet propagation and lobe morphology.

Asymmetries may also arise from differences in magnetic field structure and Faraday rotation. \citet{1983ApJ...273..128B} and \citet{1992ApJ...395..444C} found that the southern inner lobe (SIL) is more depolarized and exhibits stronger, more variable rotation measures, consistent with external Faraday depolarization caused by a clumpy ICM. In this scenario, the SIL lies behind the galactic disk, while the NIL is viewed in front. Higher-resolution VLA data and detailed rotation measure images presented by \citet{1992ApJ...395..444C} support this picture. Furthermore, \citet{1998A&ARv...8..237I} reported that the NIL is approximately 40\% brighter in radio emission than the SIL, reinforcing the idea that asymmetries may reflect environmental differences or variable energy injection.

In addition to environmental and magneto-ionic effects, several morphological features, such as the broad curvature of the northern jet, jet-counterjet asymmetry, and off-axis emission within the inner lobes, suggest a role for jet precession in shaping the large-scale structure \citep{1983PASA....5..241H, RevModPhys.56.255, 2003ApJ...594L.103G, 2019MNRAS.482..240K}. Jet precession can arise from instabilities in the accretion flow or from the presence of misaligned binary black holes \citep{2019MNRAS.482..240K}, and has been invoked to explain a variety of morphologies observed in double-lobed radio galaxies \citep{2002MNRAS.330..609D, 2014A&A...561A..30M}. Recent three-dimensional hydrodynamic simulations have demonstrated that precession can significantly influence jet morphology, dynamics, and observable features \citep{2020MNRAS.499.5765H, 2014A&A...561A..30M, 2016MNRAS.458..802N}. Quantitative morphological indicators such as curvature (C), point symmetry (S), and lobe misalignment (E) have been developed to diagnose precession signatures in synthetic radio images \citet{2020MNRAS.499.5765H}. \citet{1999MNRAS.307..750M} proposed that the misalignment observed in Centaurus~A’s jet structures could be explained by precession on a timescale of $1-2\,\mathrm{Myr}$, consistent with the morphology of the NML and its apparent connection to the inner jet.

Building on this, \citet{2019MNRAS.482..240K} estimated a precession period of $1.0\pm 0.5\,\mathrm{Myr}$ by comparing Centaurus~A with Hydra~A and using bow shock expansion timescales inferred from X-ray data. This relatively short period supports the existence of a sub-parsec-scale binary supermassive black hole at the galaxy’s center, with a separation of $\lesssim 0.05\,\mathrm{pc}$. The dynamical influence of such a system provides a plausible mechanism for long-term jet reorientation, consistent with the complex radio morphology and episodic jet activity observed in Centaurus~A. Unlike in powerful FRII galaxies, where jet reorientation may occur abruptly, the smooth curvature and gradual alignment transitions in Centaurus~A suggest a slow-precession scenario, likely driven by accretion disk instabilities or gravitational torques. Within this framework, the inner lobes offer compelling observational evidence of ongoing jet axis evolution \citep{2019MNRAS.482..240K}.

In this paper, we present three-dimensional relativistic hydrodynamic (RHD) simulations of precessing jets in Centaurus~A, performed using the \texttt{PLUTO} code \citep{2007ApJS..170..228M}. Our primary aim is to investigate how jet precession dynamically interacts with the ICM and how this interaction influences the morphology of the inner lobes. We aim to test whether jet precession can reproduce the observed morphological asymmetries in the inner lobes of Centaurus~A. To this end, we compare the resulting structures from our simulations with observed radio images. The simulations are embedded in a realistic ambient medium, modeled using density and pressure profiles derived from \textit{Chandra} X-ray observations \citep{2015MNRAS.447.1001W}. Synthetic tracer images are used to track the evolving shape of the lobes and identify characteristic features produced by precession. We further adopt the leptonic parameters from \citet{2016A&A...595A..29S} to evaluate the thermal and magnetic pressures in our simulated inner lobes, thereby connecting our dynamical models to the energetics implied by the observed high-energy emission. Additionally, we apply the curvature, symmetry, and misalignment method to quantify how jet precession imprints on the large-scale morphology of the lobes over time. This paper is organized as follows: Section~\ref{sec:setup} describes the simulation setup and physical conditions; Section~\ref{sec:simulations} presents the resulting jet morphology and dynamical evolution; and Section~\ref{sec:conclusion} summarizes the main findings.

\section{Jet Simulations}
\label{sec:setup} 

\subsection{RHD Equations}

For our simulations, we use the publicly available \texttt{PLUTO} code \citep{2007ApJS..170..228M}, version~4.4, to perform three-dimensional hydrodynamic modeling of the interaction between a precessing relativistic jet and the ICM in Centaurus~A. We employ the RHD module with a two-shock \texttt{hllc} Riemann solver and second-order Runge–Kutta (RK2) time-stepping. The equations governing the conservation of mass, momentum, and energy are \citep{1959flme.book.....L}:

\begin{equation}
\frac{\partial D}{\partial t} + \nabla \cdot (D \mathbf{v}) = 0,
\end{equation}
\begin{equation}
\frac{\partial \mathbf{M}}{\partial t}
+ \nabla\!\cdot\!\left(\mathbf{M}\otimes\mathbf{v} + p\,\mathbf{I}\right)=0,
\end{equation}
\begin{equation}
\frac{\partial E}{\partial t} + \nabla \cdot \left[ (E + p) \mathbf{v} \right] = 0,
\end{equation}
where \(D = \Gamma \rho\) is the laboratory-frame mass density, \(\mathbf{M} = \Gamma^2 \rho h \mathbf{v}\) is the momentum density, and \(E = \Gamma^2 \rho h - p\) is the total energy density in the lab frame. Here, \(\rho\) is the rest-mass density, \(\mathbf{v}\) is the three-velocity, \(p\) is the pressure, \(h = (e + p)/\rho\) is the specific enthalpy, \(e\) is the sum of the internal and rest-mass energy densities, and \(\Gamma = (1 - v^2/c^2)^{-1/2}\) is the Lorentz factor. The speed of light \(c\) is set to unity.

For the equation of state (EOS), we adopt the Taub–Matthews (TM) EOS \citep{2005ApJS..160..199M}, which expresses the specific enthalpy as
\begin{equation}
h = \frac{5}{2} \Theta + \sqrt{ \frac{9}{4} \Theta^2 + 1 },
\label{eq:tm}
\end{equation}
where \(\Theta = p / \rho\) is a temperature-like variable (in units with \(c = 1\)).

\subsection{Cluster environment}

In order to construct simulations of the inner jet propagation in Centaurus~A, we used X-ray observations from \citet{2019MNRAS.485..872W} to constrain the principal physical parameters of the ICM encountered by the jet. The authors used over $260\,\mathrm{ks}$ of \textit{Chandra} Advanced CCD Imaging Spectrometer(ACIS) X-ray data to extract temperature, density, and pressure profiles of the ICM in Centaurus~A by performing a spectral deprojection in regions adjacent to the jet.

We assume that the atmosphere of Centaurus~A is in hydrostatic equilibrium and follows a spherically symmetric density profile given by:
\begin{equation}
\rho(r)=\rho_0\left[1+\frac{r^2+r_s^2}{r_c^2}\right]^{-3\beta/2},
\end{equation}
where \( r_c = 1.0\,\mathrm{kpc} \) is the core radius, \( r_s = 0.1\,\mathrm{kpc} \) is a smoothing parameter introduced to maintain consistency between our model and the observational constraints by ensuring proper normalization of the density profile. \(\rho_0\) is the central density at \(r = 0\) (set equal to the unit density used in the simulation), and \(\beta = 0.5\) controls the slope of the density profile. The central density \(\rho_0\) is scaled to physical units by requiring that the profile matches the observed ambient electron density at the jet injection radius (\(r \approx 0.5\,\mathrm{kpc}\))~\citep{2019MNRAS.485..872W}.

The intracluster medium is modeled as a hydrostatic atmosphere with a temperature of $0.65\,\mathrm{keV}$, consistent with deprojected radial temperature profiles, and an external pressure of \(5.72\times 10^{-11}\,\mathrm{dyne\,cm^{-2}}\). The ambient electron density at the jet base is approximately \(9.156\times 10^{-26}\,\mathrm{g\,cm^{-3}}\). The external sound speed is taken to be $299.8\,\mathrm{km\,s^{-1}}$, placing the simulated jet in a supersonic regime relative to the environment \citep{2019MNRAS.485..872W}. These conditions provide a physically motivated context for analyzing the propagation and disruption of a precessing relativistic jet in a low-power radio galaxy, enabling comparison with both the inner kpc-scale structure and the extended lobes of Centaurus~A.

\subsection{3D hydrodynamic model}

Our simulations are based on a 3D RHD model inspired by the jet injection framework described by \citet{2014A&A...561A..30M, 2020MNRAS.499.5765H}. The jet is injected through a spherical boundary region of radius \( r_0 = 0.5\,\mathrm{kpc} \), located at the center of the computational domain. This region acts as a controlled injection zone where physical quantities are overwritten at each time step to represent a precessing bipolar jet.

\begin{figure}
   \centering
   \includegraphics[width=0.4\textwidth]{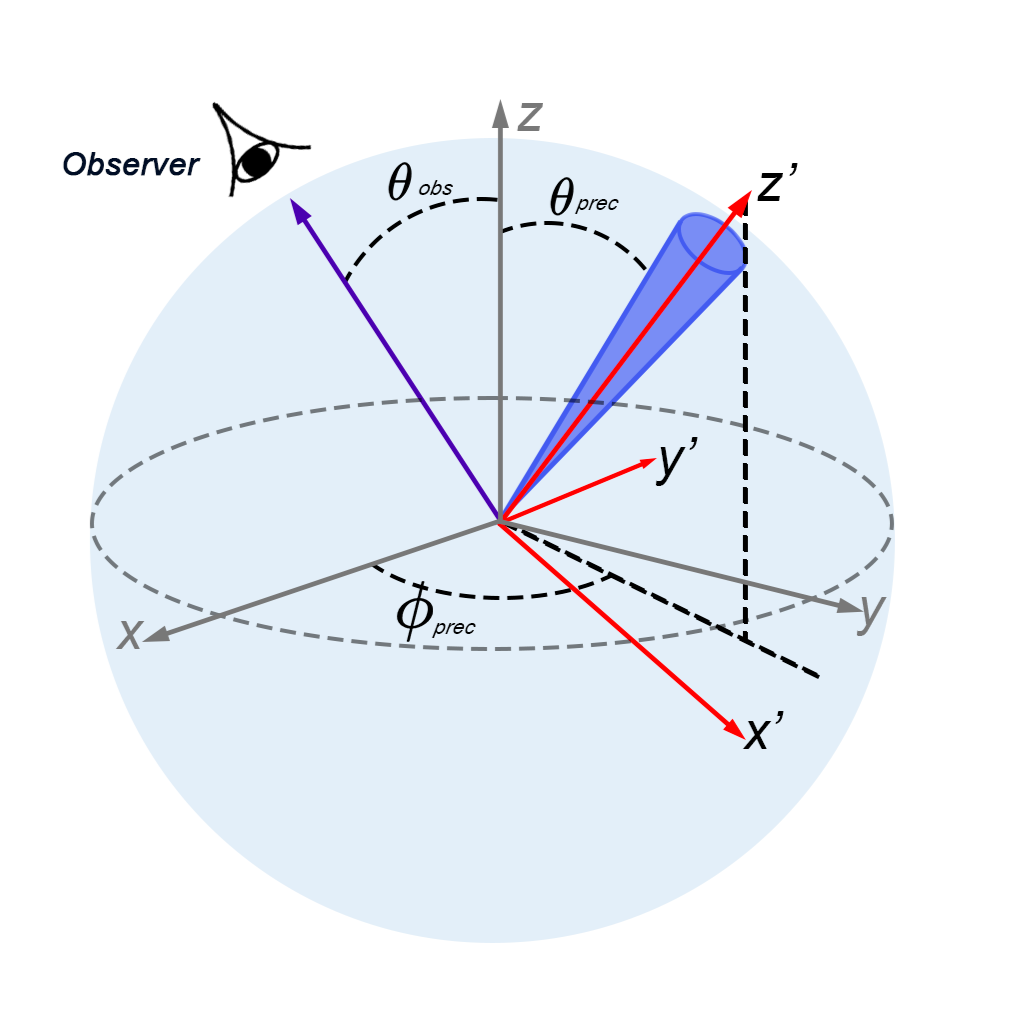}
   \caption{Schematic representation of the jet precession geometry and viewing configuration. The jet precesses around the $z$-axis with opening angle $\theta_{\rm prec}$ and azimuthal angle $\phi_{\rm prec}$. The observer viewing angle $\theta_{\rm obs}$ sets the orientation for synthetic observations.}
   \label{fig:geometry}
\end{figure}

Within this sphere, we define two opposing cones corresponding to the jet and counterjet, each with a fixed half-opening angle \( \alpha_j \). The orientation of the cone axis evolves according to the precession parameters, including the precession angle \( \theta_{\rm prec} \), the precession period \( P_{\rm prec} \), and the azimuthal angle of precession \( \phi_{\rm prec} = 2\pi t / P_{\mathrm{prec}} \). The instantaneous jet direction is computed at each time step, describing a circular motion around the precession axis, as seen in Figure \ref{fig:geometry}.

The precession angle is \( \theta_{\rm prec} = 50^\circ \) \citep{2015MNRAS.447.1001W} with respect to the \( z \)-axis. By adjusting the orientation of the precession axis, we ensure that the synthetic morphology from our simulations closely reproduces the observed radio features of the inner lobes. The radius of the jet at injection is computed from the intersection of the conical jet surface with the spherical boundary, given by \( r_j = r_0 \tan \alpha_j \). The velocity field within the jet cones is purely radial and aligned with the local jet axis. This setup allows us to capture the interaction between the precessing jet and the galactic environment, providing an understanding of the formation of the S-shaped inner lobes.

The simulations are performed in Cartesian coordinates using a computational grid composed of \( 600 \times 600 \times 600 \) cells. The central region, spanning from \( -5 \) to \( 5 \)~$\mathrm{kpc}$ in all three spatial directions, is covered by a uniform grid of \( 400 \times 400 \times 400 \) cells, providing high spatial resolution near the jet base. To extend the domain and reduce boundary effects, this uniform core is flanked by two stretched zones on each side: 100 cells from \( -12 \) to \( -5 \)~$\mathrm{kpc}$ and from \( 5 \) to \( 12 \)~$\mathrm{kpc}$ in both the \( x \) and \( y \) directions, and from \( -15 \) to \( -5 \)~$\mathrm{kpc}$ and from \( 5 \) to \( 15 \)~$\mathrm{kpc}$ in the \( z \) direction.

\subsection{Jet parameters and initial conditions}
\label{sec:jetparams}

The precessing jet model for Centaurus~A adopts physically motivated parameters guided by both observational constraints and analytic studies \citep{2019MNRAS.485..872W}. The jet is injected with a bulk velocity of \( 0.667c \) (corresponding to a Lorentz factor \( \Gamma = 1.34 \)), a value suitable for a trans-relativistic plasma of relativistic electrons and sub-relativistic protons. The jet's internal Mach number is approximately \( \mathcal{M} \sim 5.5 \), and the half-opening angle is set to \( 12^\circ \).

The surrounding ICM is treated as a hydrostatic, spherically symmetric atmosphere with a temperature of \( kT = 0.65 \)~keV, external pressure \( P = 5.72 \times 10^{-11}~\mathrm{dyne\,cm^{-2}} \), and a sound speed of approximately \( 300 \)~km\,s\(^{-1} \) \citep{2003ApJ...592..129K}. At injection, the jet has a rest-mass density of \( \rho_j = 9.156 \times 10^{-29}~\mathrm{g\,cm^{-3}} \) and pressure \( p_j = 8.19 \times 10^{-10}~\mathrm{dyne\,cm^{-2}} \). Assuming a uniform top-hat jet profile, the total kinetic power is given by:
\begin{equation}
Q_j = \pi r_j^2 v_j \left[ \Gamma_j (\Gamma_j - 1) \rho_j c^2 + \frac{\gamma}{\gamma - 1} \Gamma_j^2 P_j \right],
\label{eq:Qj}
\end{equation}
where \( \gamma = 13/9 \) is the polytropic index, \( r_j \) the jet radius, and \( v_j \) is the bulk flow speed. The value $\gamma = 13/9$ is the effective adiabatic index of a charge-neutral plasma with ultra-relativistic electrons and sub-relativistic protons, consistent with the trans-relativistic jet composition adopted in Section~\ref{sec:jetparams} and with the Taub--Matthews EOS (Eq.~\ref{eq:tm}) in the same regime \citep{2005ApJS..160..199M}. We follow \citet{2019MNRAS.485..872W}, who justifies this value for the Centaurus A jet based on its effective temperature range ($0.5~\mathrm{MeV} \lesssim kT \lesssim 1~\mathrm{GeV}$). Using the parameters listed in Table~\ref{tab:jetparams}, this yields a jet power of approximately $10^{43}~\mathrm{erg\,s^{-1}}$, consistent with estimates for the kinetic luminosity of the inner jet in Centaurus~A \citep{2019MNRAS.485..872W}. A complete summary of the jet and environmental parameters extracted from observations and used in the simulation is presented in Table~\ref{tab:jetparams}.

\begin{table}
\begin{center}
\caption{Hydrodynamic setup of the precessing jet model for Centaurus~A \citep{2019MNRAS.485..872W}.}
\begin{tabular}{lccc}
\hline
\textbf{Fluid} & \textbf{Quantity} & \textbf{Value} & \textbf{Units} \\
\hline
\textbf{Jet} \\
& Density & \( 9.156 \times 10^{-29} \) & \( \mathrm{g\,cm^{-3}} \) \\
& Velocity & \( 0.667c \) & \( \mathrm{cm\,s^{-1}} \) \\
& Lorentz factor & 1.34 & -- \\
& Internal sound speed & \( 0.12c \) & \( \mathrm{cm\,s^{-1}} \) \\
& Mach number & 5.5 & -- \\
& Pressure & \( 8.19 \times 10^{-10} \) & \( \mathrm{dyne\,cm^{-2}} \) \\
& Opening angle (\( \alpha_j \)) & \( 12^{\circ} \) & -- \\
& Precession period (\( P \)) & 1.8 & \( \mathrm{Myr} \) \\
& Jet power & \( 10^{43} \) & \( \mathrm{erg\,s^{-1}} \) \\
\textbf{Environment} \\
& External sound speed & 299.8 & \( \mathrm{km\,s^{-1}} \) \\
& Density & \( 9.156 \times 10^{-26} \) & \( \mathrm{g\,cm^{-3}} \) \\
& ICM gas temperature & 0.65 & \( \mathrm{keV} \) \\
& Pressure & \( 5.72 \times 10^{-11} \) & \( \mathrm{dyne\,cm^{-2}} \) \\
\hline
\end{tabular}
\label{tab:jetparams}
\end{center}
\end{table}

\subsection{Synthetic radio images}

We generated synthetic synchrotron-intensity images from the simulation output to assess the observable impact of jet precession on the morphology of Centaurus~A. Following the general method of \citet{2007ApJS..173...37S}, we compute the synchrotron rest-frame emissivity as
\[
j_\nu = \lambda\, p^{\alpha + 3/2}\, \delta^{2 + \alpha},
\]
where \(p\) is the gas pressure (assumed to be proportional to the total particle pressure), \(\lambda\) is a passive scalar tracer that distinguishes jet plasma from the ambient medium, and \(\delta\) is the Doppler factor accounting for relativistic beaming effects. We adopt a spectral index of \(\alpha = 1.0\) and assume that the magnetic pressure scales with the gas pressure. 

To simulate different viewing orientations, we apply two sequential rotations to the simulation data cube following \citet{2016MNRAS.458..802N}: an azimuthal rotation by angle $\chi$ about the $y$-axis, which rotates the observer around the jet precession axis, and an inclination rotation by angle $\theta_{\rm obs}$ about the $x$-axis, which sets the viewing angle with respect to the precession axis. Here, $x$, $y$, and $z$ refer to the fixed simulation coordinate system, where the $z$-axis is aligned with the jet precession axis. The Doppler factor is computed locally in each cell as
\[
\delta = \frac{1}{\Gamma \bigl(1 - \beta \cos\theta'\bigr)},
\]
where \(\Gamma\) is the Lorentz factor, \(\beta\) is the flow velocity in units of \(c\), and \(\theta' = \cos^{-1}\bigl(v'_{\rm LOS}/\lvert\mathbf{v}'\rvert\bigr)\) is the angle between the cell's velocity vector and the line of sight in the rotated frame, with \(v'_{\rm LOS}\) denoting the line-of-sight component of the rotated velocity vector \(\mathbf{v}'\).

After these rotations, we integrate the emissivity along the line of sight to produce synthetic emission maps:
\[
I_\nu = \int j_\nu\,\mathrm{d}s,
\]
where the integration is performed along one of the axes of the rotated data cube. We explored several combinations of $\theta_{\rm obs}$ and $\chi$ to identify the orientation that best matches the observed large-scale structure of Centaurus~A.

Among the tested combinations, we found that $\chi = 0^\circ$ and $\theta_{\rm obs} = 210^\circ$ yields a synthetic morphology that best reproduces the observed S-shaped structure and north--south asymmetry of Centaurus~A's inner lobes (Figure~\ref{fig:cenA_inner_lobes}). This orientation produces an $x$--$z$ sky projection (in the original simulation frame), where the integration is performed along the line of sight through the rotated cube. Within this projection, the jet precesses with angle $\theta_{\rm prec}=50^\circ$ about the $z$-axis, naturally producing the characteristic curvature and lobe asymmetry observed in Centaurus~A.

Each resulting synthetic map was convolved with a Gaussian kernel to approximate the effect of finite instrumental resolution. This approach allows us to assess how the observed morphology depends on both the viewing orientation and the projection geometry.

\section{Results}
\label{sec:simulations}

In this section, we present the simulation results for Centaurus~A. We first explore how varying the precession period affects the resulting jet and lobe morphologies. We then discuss the underlying three-dimensional simulation dynamics that give rise to these features, followed by an analysis of specific morphological markers of jet precession and leptonic emission. These measures allow us to quantify the influence of jet precession on the observed S-shaped structure of Centaurus~A and assess how well the simulations reproduce the asymmetries and deviations from linearity observed in the inner lobes.

\subsection{Morphology and jet structure}
\label{sec:morphology}

We present the selection process for the precession parameters adopted in our model, guided by comparisons between synthetic radio emission images and observations of Centaurus~A. In particular, we focus on determining the precession period and jet orientation that best reproduces the morphology of the inner lobes. Observational studies estimate the age of these lobes to be approximately $2\,\mathrm{Myr}$ \citep{2009MNRAS.395.1999C, 2013A&A...558A..19W, 2015MNRAS.447.1001W}, consistent with a jet precession timescale in the range of $1-2\,\mathrm{Myr}$ \citep{2019MNRAS.482..240K}. We test a range of precession periods and viewing angles, and identify the configuration that most closely matches the observed S-shaped radio morphology.

\begin{figure*}
   \centering
    \includegraphics[angle=0,width=1.0\textwidth]{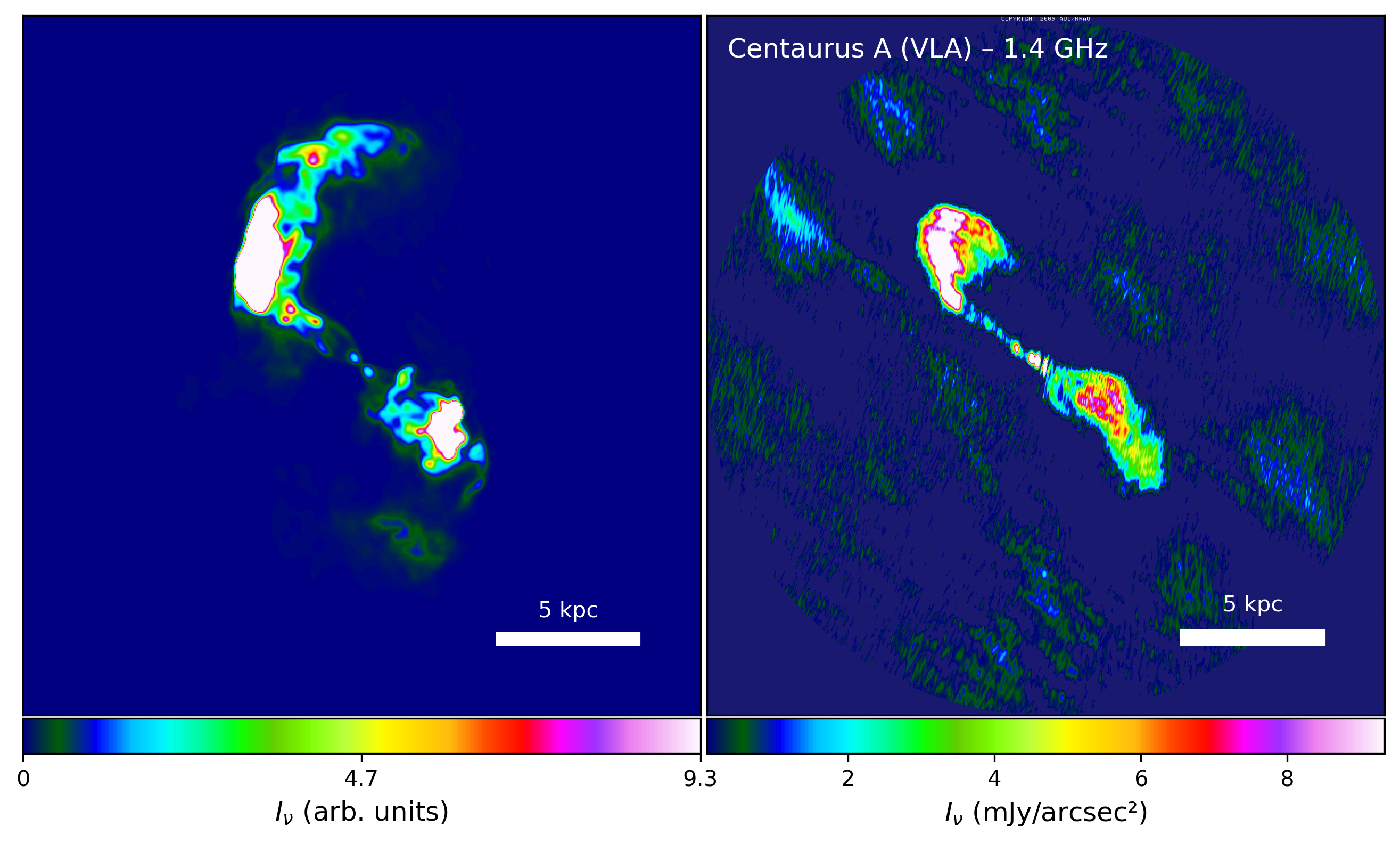}
    \caption{Comparison between synthetic radio emission from the simulation (left) and VLA observations of Centaurus~A at $1.4\,\mathrm{GHz}$ (right). The simulation includes a jet precessing with an angle of $\theta_{\rm prec} = 50^\circ$ around the $z$-axis. Color bars indicate the intensity scale in arbitrary units (left) and mJy/arcsec$^2$ (right). Scale bars correspond to 5 kpc in both panels.}
   \label{fig:compare_sim_obs}
\end{figure*}

Figure~\ref{fig:compare_sim_obs} displays the resulting images for period $P=1.8\,\mathrm{Myr}$, normalized consistently for direct morphological comparison. The observed inner lobes of Centaurus~A, as revealed by high-resolution VLA observations at 1.4~GHz \citep{2009MNRAS.395.1999C}, exhibit broad, laterally extended emission regions with a pronounced S-shaped symmetry. The radio emission is limb-brightened, and the lobes connect smoothly back to the central nucleus via moderately curved jet channels. The morphology closely resembles the observed structure, particularly in the distribution of bright emission regions and the overall lobe curvature. The simulated lobes show lateral expansion and asymmetry, including the offset between the northern and southern lobes, which mirrors features in the VLA image. The match between the $P=1.8\,\mathrm{Myr}$ simulation and the observed morphology thus supports the interpretation that the inner lobe structure of Centaurus~A is shaped by a precessing jet with a period on the order of $2.0\,\mathrm{Myr}$, consistent with previous age estimates of the inner lobes \citep{2013A&A...558A..19W, 2015MNRAS.447.1001W}.

The brightest observed regions (in white and red) lie along the inner northern and southern lobes, near the jet axis. These are likely sites of synchrotron and IC emission driven by relativistic particles and magnetic fields, associated with internal shocks, re-confinement, or interactions with dense ambient gas \citep{2002ApJ...569...54K, 2009MNRAS.395.1999C, 2003ApJ...593..169H}. The simulation reproduces this structure, with bright core emission transitioning into curved lobes that broaden laterally. Bright simulated regions align with high-pressure shocks, consistent with observed radio and X-ray correlations. The precession and tilt produce a brighter northern lobe and an off-axis displacement of emission maxima, reflecting the observed asymmetry in Centaurus~A’s inner lobes and supporting interpretations of precessional motion or environmental deflection \citep{1992ApJ...395..444C, 2019MNRAS.482..240K}. The limb-brightened features and bright knots in the synthetic image correspond to internal shock regions visible in the pressure slices of Figure~\ref{fig:3D_pressure_density}.

However, despite the broad morphological agreement, some differences are evident. The observed image shows a more filamentary and patchy structure, with clear knotty substructures along the jet axis, particularly in the southern lobe \citep{2003ApJ...592..129K, 2024ApJ...974..307B, 2025MNRAS.539.3697D}. These may reflect complex physical processes not fully captured in the current hydrodynamic simulation, such as magnetic field dynamics, which play a crucial role in confining and accelerating particles within astrophysical jets and lobes, interactions with the intracluster medium and density clumps~\citep{1988Natur.335..146N, 2015ApJ...802...87N}, or intermittent jet activity~\citep{2025MNRAS.539.3697D}. Additionally, the radio brightness in the simulation is smoother and more symmetric, possibly due to the absence of turbulent instabilities or resolution limitations. While the comparison validates the idea that a precessing jet offers a plausible explanation for Centaurus~A's observed emission morphology, a more comprehensive model that includes magnetic fields and interactions with denser clumps or filaments in the ICM could further improve the simulation’s reproduction of the inner lobe curvature and brightness.

\subsection{Jet-ICM interaction and dynamics}

\begin{figure*}
   \centering
    {\includegraphics[angle=0,width=1.0\textwidth]{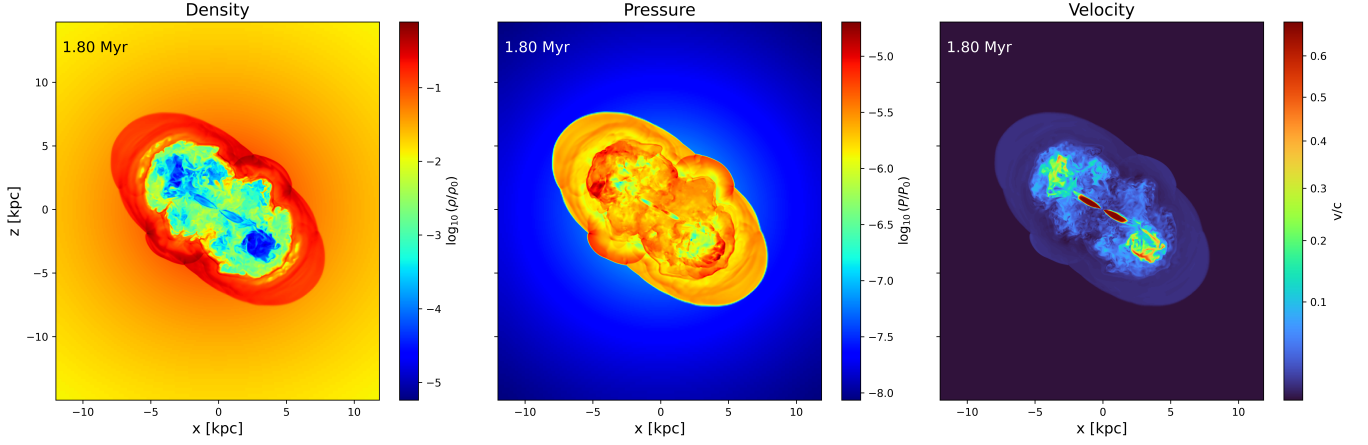}}
\caption{Slices from the 3D precessing jet simulation at \(t=1.80~\mathrm{Myr}\). Each panel shows a mid\textendash{}plane slice through the \(x\)–\(z\) plane centered on the jet axis: \emph{left} \(\log_{10}(\rho/\rho_{0})\), \emph{middle} \(\log_{10}(P/P_{0})\), and \emph{right} the velocity magnitude \(v/c\). The images show the over\textendash{}pressured cocoon, bow shocks, and the fast jet spine. Normalizations are \(\rho_{0}=9.156\times10^{-26}\,\mathrm{g\,cm^{-3}}\) and \(P_{0}=5.72\times10^{-11}\,\mathrm{dyne\,cm^{-2}}\).}
    \label{fig:3D_pressure_density}
\end{figure*}

Figure~\ref{fig:3D_pressure_density} shows mid–plane slices of the rest–mass density, gas pressure, and velocity magnitude in the \(x\)–\(z\) plane at the simulation time \(t=1.80~\mathrm{Myr}\). The \emph{left}, \emph{middle}, and \emph{right} panels display \(\log_{10}(\rho/\rho_{0})\), \(\log_{10}(P/P_{0})\), and \(v/c\), respectively. The panels illustrate how precession and hydrodynamic interaction with the ambient medium shape the structure of the cocoon and inner lobes. Both the density and pressure images reveal a developed bipolar outflow structure. Multiple internal shocks are visible along the jet beam, with localized pressure enhancements and filamentary density structures within the cocoon. The jet terminates in a broad, asymmetric bow shock, indicative of strong interaction with a stratified ambient medium~\citep{1982A&A...113..285N, 1983ApJ...273..128B}. The surrounding cocoon displays a layered pressure structure, with higher pressure near the jet axis and diffuse, lower–pressure material at the edges. These features resemble the S–shaped distortion observed in the radio lobes of Centaurus~A~\citep{1983ApJ...273..128B, 1998A&ARv...8..237I}, suggesting that the morphology of the inner lobes could be dynamically driven by jet precession and interaction with the intracluster medium. The pronounced pressure contrast between the cocoon and the undisturbed ICM reflects a shock–driven expansion process, as predicted in theoretical models~\citep{RevModPhys.56.255, 1982A&A...113..285N} and supported by observations of X–ray–emitting shells in Centaurus~A~\citep{2003ApJ...592..129K, 2009MNRAS.395.1999C}. The velocity panel illustrates how jet precession and interaction with the ambient medium modify the jet’s velocity structure, both radially and laterally; the fast spine reaches \(v/c \gtrsim 0.6\) and decelerates into a broader, asymmetric cocoon. This morphology reflects the effects of sustained precession and environmental interaction, resulting in momentum redistribution and entrainment, in accordance with~\citet{2020MNRAS.499.5765H} and~\citet{2016MNRAS.458..802N}.

\begin{figure*}
   \centering
    \includegraphics[angle=0,width=1.0\textwidth]{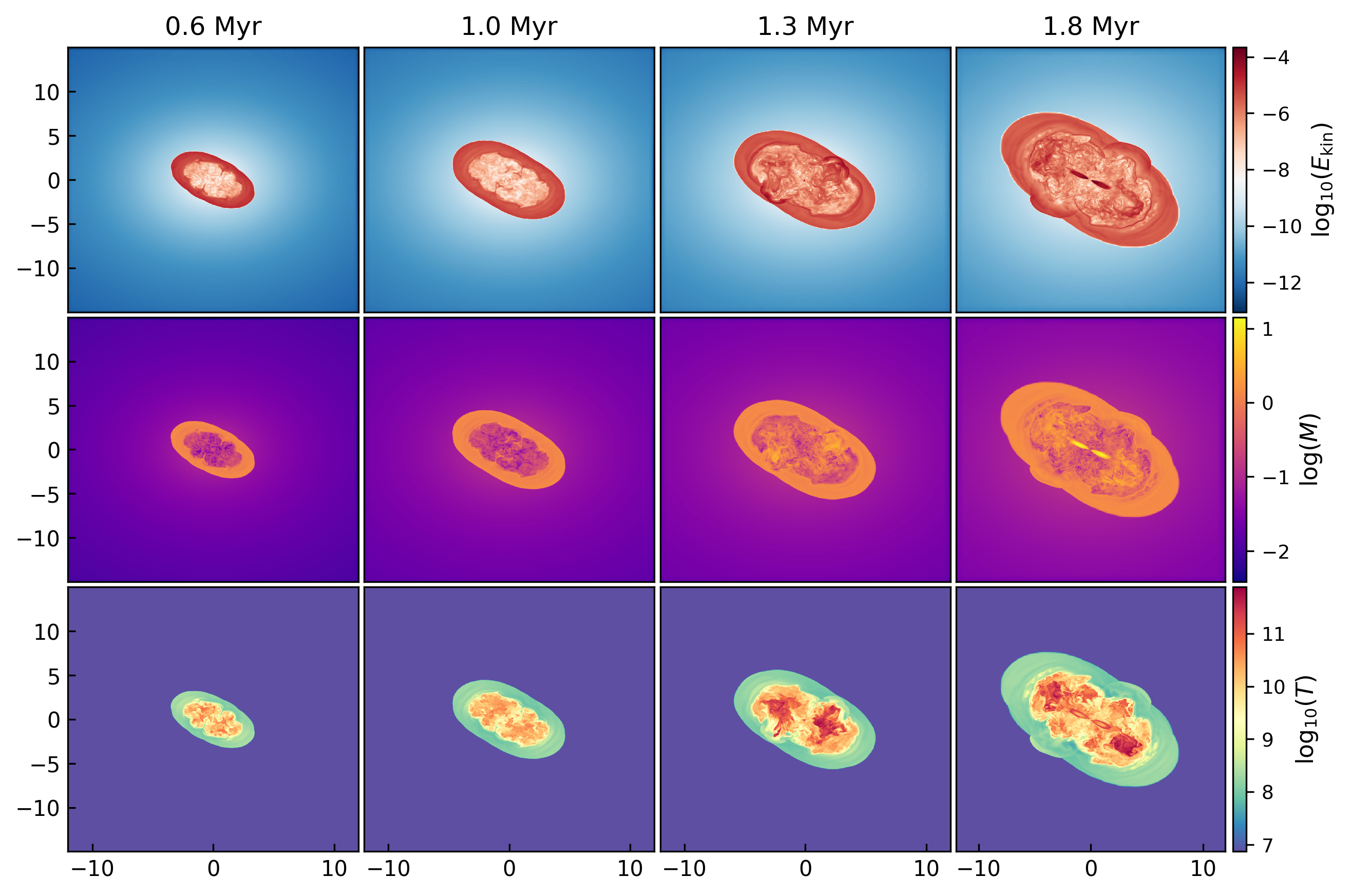}
    \caption{Slices in the \( x \)--\( z \) plane of kinetic energy density, Mach number, and bulk thermal gas temperature assuming a hydrogen plasm} for the 3D precessing jet simulation of Centaurus~A. Each column shows a snapshot at a different simulation time (0.6, 1.0, 1.3, and $1.8\,\mathrm{Myr}$), covering the full precession period. The spatial domain spans $[-12, +12]$ $\mathrm{kpc}$ along both axes, centered on the jet injection region. The images capture the dynamical evolution of the jet-driven cocoon, showing the formation of internal shocks, bow shock propagation, and thermalization of the ambient medium. The kinetic energy density is shown in simulation units (\( \rho_0 c^2 \)), and the temperature is expressed in $\mathrm{kelvin}$.
    \label{fig:evolution_energy_M_T}
\end{figure*}

Figure~\ref{fig:evolution_energy_M_T} presents logarithmic slices of the kinetic energy density, Mach number, and temperature (assuming a hydrogen plasma) along the \( x \)--\( z \) plane at various stages of jet evolution, spanning up to one full precession period ($1.8\,\mathrm{Myr}$). In the top row, the kinetic energy density highlights the jet beam and shock fronts. At early times, the energy is concentrated along a narrow, well-collimated spine. As precession progresses, this energy spreads laterally, indicating increasing entrainment and turbulent mixing with the ICM, consistent with jet broadening observed in the inner lobes of Centaurus~A~\citep{1992ApJ...395..444C, 1983ApJ...273..128B}. The flaring of the kinetic energy contours resembles features seen in observed X-ray knots and flares along the jet path~\citep{2002ApJ...569...54K}.

The middle row shows the Mach number distribution. Initially, the jet maintains a high internal Mach number \( \sim 5.0 - 6.0\), in agreement with expectations from analytical models of flow~\citep{2015MNRAS.447.1001W}. However, as the jet interacts with the surrounding ICM, the Mach number locally decreases, particularly near shock fronts and in turbulent shear regions. This behavior is consistent with theoretical predictions that entrainment reduces the Mach number and induces subsonic regions near internal working surfaces and cocoon boundaries~\citep{2014MNRAS.441.1488P}. Specifically for Centaurus~A, \citet{2015MNRAS.447.1001W} estimated sonic Mach numbers in the range \( \sim 3.6 - 5.3\), consistent with a jet that starts out supersonic but slows down due to interaction with the surrounding medium.

The bottom row displays the bulk thermal gas temperature, calculated from the specific enthalpy in Eq.~\ref{eq:tm}, assuming a hydrogen plasma composition. Interactions between the jet and the ICM drive shock heating, elevating the thermal temperature in a shell surrounding the jet beam. While the jet interior remains hot due to relativistic injection conditions, the surrounding cocoon exhibits complex temperature gradients, reflecting repeated shock crossings and turbulent dissipation. Note that this thermal temperature characterizes the bulk plasma pressure support and is distinct from the energy distribution of the nonthermal and synchrotron-emitting electron population. Observationally, similar thermal structures have been inferred from X-ray shells and diffuse soft X-ray emission around the southwestern lobe of Centaurus~A~\citep{2003ApJ...592..129K, 2009MNRAS.395.1999C}. \citet{2015MNRAS.447.1001W} reported that the temperature along the inner jets of Centaurus~A varies between approximately $8.6$ and $4.7\,\mathrm{keV}$, reflecting the thermal structure shaped by jet propagation and interaction with the ambient medium.

\subsection{Precession Signatures in the Jet}
\label{sec:precession_markers}

The morphological evidence for jet precession in radio galaxies is often subtle and encoded in large-scale structures. The presence of jet precession in the inner lobes of Centaurus~A can be inferred through three key morphological markers: curvature (\(C\)), point symmetry (\(S\)), and lobe axis misalignment (\(E\)). These indicators, defined by \citet{2019MNRAS.482..240K} and adopted by \citet{2020MNRAS.499.5765H}, provide a quantitative framework for identifying precession in relativistic jets. Following this methodology, we describe the jet/counterjet structure and the morphology of the inner lobes in Centaurus~A.

\begin{figure*}
   \centering
   \includegraphics[angle=0,width=0.8\textwidth]{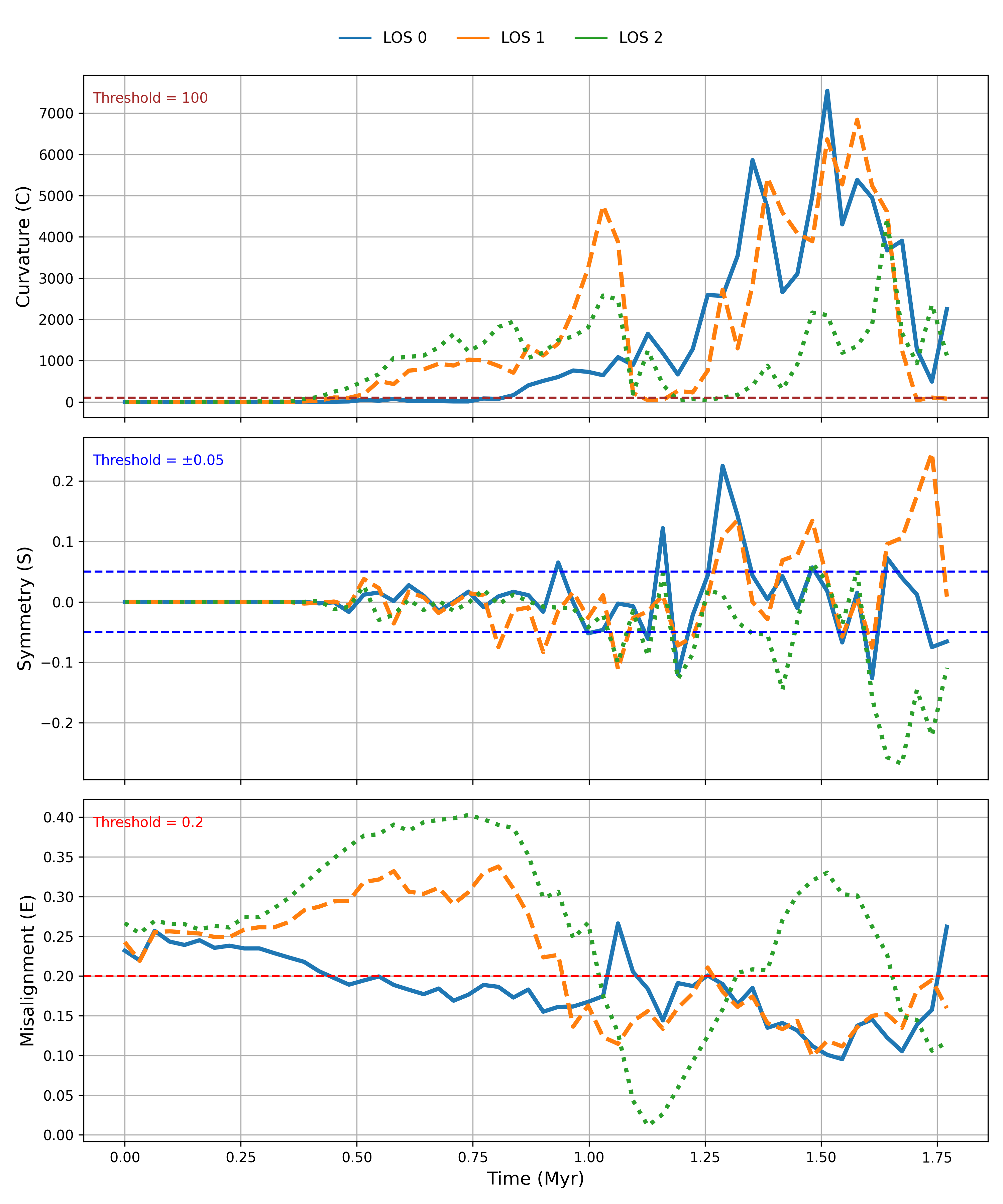} 
     \caption{Time evolution of jet morphology indicators for the precessing jet model, shown for three LOS (LOS 0 - \(x\), LOS 1 - \(y\), and LOS 2 - \(z\)). Top: Curvature ($C$), computed as the mean squared deviation of the jet spine from a linear fit in pixel units. The dashed red line marks the adopted threshold $C = 100$, above which the jet is considered significantly bent. Middle: Symmetry ($S$), indicating the balance between the northern and southern lobes. Values near zero reflect symmetric jet/counterjet structure, while departures from the dashed blue lines at $S = \pm 0.05$ denote lobe dominance. Bottom: Misalignment ($E$), quantifying the lateral offset of the jet spine from the central axis, normalized by the lobe width. The red dashed line at $E = 0.2$ separates misaligned phases from baseline values. 
    }
    \label{fig:markers}
\end{figure*}

The jet curvature (\(C\)) is quantified by fitting a straight line to its projected path and calculating the deviation from linearity. The curvature index is defined as:
\begin{equation}
    C = \frac{1}{N} \sum_{i=1}^{N} \left[ y_i - (a x_i + b) \right]^2,
\end{equation}
where \(N\) is the number of data points along the jet, \((x_i, y_i)\) are the coordinates of the jet path, and \(a\), \(b\) are the parameters of the best-fit straight line. A high value of \(C\) indicates significant curvature, reflecting deviation from ballistic behavior. For Centaurus~A, the inner jet and counterjet exhibit clear curvature, suggesting precession-driven reorientation. Figure~\ref{fig:markers} presents the time evolution of the three key jet morphology indicators: curvature, point symmetry, and misalignment, extracted from our precessing jet simulation of Centaurus~A. These indicators are computed across three orthogonal LOS, namely along the \(x\)-, \(y\)-, and \(z\)-axes (denoted as LOS 0, 1, and 2, respectively), to assess projection effects and orientation dependence. The top panel of Figure~\ref{fig:markers} shows the evolution of jet curvature. A threshold of \(C = 100\) (indicated by the dashed red line) is used to demarcate significant curvature from near-ballistic trajectories. During early stages ($t \lesssim 0.5\ \mathrm{Myr}$), all views display low curvature as the jet is still propagating quasi-linearly. However, curvature increases rapidly in LOS 1 and 2 as precession-induced deflections accumulate, reaching peak values near the precession maximum ($\sim 1.8\ \mathrm{Myr}$). The variation across lines of sight highlights the importance of projection: while LOS 2 (aligned with the precession axis) shows the weakest deviation, LOS 0 and 1 capture the full lateral motion of the jet and are more sensitive to its bending. Our curvature threshold (\(C = 100\)) is higher than the value (\(C = 10\)) adopted by \citet{2020MNRAS.499.5765H}, primarily due to the jet’s larger inclination (\(50^\circ\)). This configuration amplifies the apparent curvature in LOS projections, necessitating an increased threshold for accurate identification. Additionally, Centaurus~A’s dense and stratified ICM \citep{2019MNRAS.485..872W} further enhances jet bending compared to the uniform medium conditions explored by \citet{2020MNRAS.499.5765H}, naturally increasing the measured curvature values.

The point symmetry (\(S\)), or rotational symmetry, between the jet and counterjet is another signature of precession. Following \citet{2019MNRAS.482..240K} and \citet{2020MNRAS.499.5765H}, the symmetry index is calculated as:
\begin{equation}
    S = \sqrt{\frac{\sum \left( d_{i,\text{jet}} \cdot (-d_{i,\text{counterjet}}) \right)}{N w^2}},
\end{equation}
where \(d_{i,\text{jet}}\) and \(d_{i,\text{counterjet}}\) are the displacements of the jet and counterjet from the lobe axis at position \(x_i\), and \(w\) is the mean lobe width. The square root in this context denotes a \textit{signed} operation: it returns the square root of the absolute value of the numerator while preserving its sign. Consequently, a positive \(S\) (\(>0\)) indicates S-shaped (point) symmetry, whereas a negative \(S\) (\(<0\)) signifies mirror symmetry between the jet and counterjet. In Centaurus~A, the inner lobes exhibit such S-symmetry, further supporting the precession scenario. The middle panel of Figure~\ref{fig:markers} displays the \(S\)-symmetry index. A value near zero denotes perfect mirror symmetry, while departures beyond \(|S| > 0.05\) (dashed blue lines) indicate dominance or deflection of one jet over the other. The \(S\) parameter fluctuates moderately across all LOS, with excursions beyond the threshold particularly pronounced in LOS 1. This behavior is consistent with precession-driven deviations that break bilateral symmetry, as also observed in the synthetic radio images (see Section~\ref{sec:morphology}).

Finally, the lobe axis misalignment (\(E\)) quantifies the average offset of the jet from its lobe center. The misalignment index is defined as:
\begin{equation}
    E = \frac{1}{N} \sum_{i=1}^{N} \frac{|d_{i,\text{jet}}|}{w/2},
\end{equation}
where \(d_{i,\text{jet}}\) is the distance of the jet from the lobe center, and \(w\) is the lobe width. An \(E\) value close to 1 indicates the jet lies near the lobe edge, a common feature in precessing systems. The inner lobes of Centaurus~A show such misalignment, with the jet often displaced from the lobe center. Figure~\ref{fig:markers} (bottom panel) shows the misalignment index. A value of \(E = 0\) corresponds to a centrally aligned jet, whereas \(E \geq 0.2\) (dashed line) indicates a significant offset and is considered indicative of precession~\citep{2020MNRAS.499.5765H}. Similar to \(C\), the \(E\) index increases with time, peaking near the end of the simulation, particularly in LOS 1 and 2. This behavior suggests increasing misalignment resulting from both the widening precession cone and the cumulative displacement of the jet trajectory.

Together, these indicators demonstrate that the inner lobes of the Centaurus~A jet exhibit all three primary morphological signatures of jet precession. These results are consistent with the interpretations proposed by \citet{2019MNRAS.482..240K} and \citet{1999MNRAS.307..750M}, supporting the hypothesis that Centaurus~A's intricate inner lobe morphology arises from gradual, sustained jet precession, likely driven by either a binary supermassive black hole system or torques originating from the accretion disk.

\subsection{Leptonic emission from the Inner Lobes}

The precessing jet model (Sections \ref{sec:morphology} - \ref{sec:precession_markers}) influences the spatial distribution of particle acceleration sites, with shocks forming at the jet-ICM interaction regions. These shocks are particularly evident in the limb-brightened regions of the inner lobes (Figure~\ref{fig:compare_sim_obs}), where our simulations show pressure enhancements of $\sim$100$\times$ the ambient value (Figure~\ref{fig:3D_pressure_density}). The study of synchrotron and inverse Compton emission from the inner lobes of Centaurus~A provides constraints on the physical conditions and particle acceleration processes in these regions (see, e.g., ~\citet{2009MNRAS.395.1999C}; \citet{2016A&A...595A..29S}; \citet{2019MNRAS.483.3444P}; \citet{2020Natur.582..356H}; \citet{2021MNRAS.505.1334W};  \citet{2025MNRAS.539.3697D}). By analyzing the spectral energy distribution (SED), we determine key parameters such as the magnetic field strength and the energy spectrum of relativistic electrons, which are essential for understanding energy dissipation and the interaction between the jet and its environment.

Early X-ray observations revealed non-thermal synchrotron emission from the inner lobes, where a \( \mathcal{M} \sim 8.0 \) shock front accelerates electrons to tens of TeV. This emission dominates over thermal components, resembling processes seen in young supernova remnants~\citep{2009MNRAS.395.1999C}. Meanwhile, compact knots along the jet, such as AX1A and AX2 knots, act as additional acceleration sites, where interactions with stellar winds produce stationary bow shocks and boost electrons to PeV energies~\citep{2025MNRAS.539.3697D}. These features are corroborated by H.\,E.\,S.\,S. observations of TeV gamma rays, confirming IC scattering of infrared and CMB photons by ultra-relativistic electrons~\citep{2020Natur.582..356H}.

The lobes also serve as reservoirs for cosmic rays, with proton populations potentially accelerated to $>10^{20}$\,eV over their $\sim 500\,\mathrm{Myr}$ lifetime~\citep{2009MNRAS.395.1999C}. Recent \textit{Fermi}-LAT observations of high-directional-quality gamma rays ($6.5-300\,\mathrm{GeV}$) reveal emission elongated along the inner lobes, with a spatial distribution coinciding with the radio structures~\citep{2019MNRAS.483.3444P}. This suggests that a significant fraction of the gamma-ray core's emission above the break energy ($\sim 2.8\,\mathrm{GeV}$) originates in the lobes, likely through IC scattering of CMB photons by high-energy electrons (Lorentz factors $\sim 2.5 \times 10^6$) or hadronic interactions in dense filamentary substructures~\citep{2019MNRAS.483.3444P}. The gamma-ray flux asymmetry between the northern and southern lobes aligns with Faraday depolarization observations, hinting at environmental differences influencing particle acceleration~\citep{1983ApJ...273..128B,2019MNRAS.483.3444P}.

We analyzed the N3 region of Centaurus~A's northern lobe, as defined by~\citet{2016A&A...595A..29S}, which encompasses the inner portions of the lobe. Using relativistic RHD simulations discussed in previous sections and the \textsc{naima} package~\citep{naima}, we reproduced the SED of this region, finding agreement with the volume-averaged magnetic field strength of $B=1.8\,\mu\mathrm{G}$, derived from synchrotron and IC modeling in~\citet{2016A&A...595A..29S}. The electron population in this region was characterized by a power-law spectrum with an exponential cutoff:
\begin{equation}
N(E) = A \left(\frac{E}{E_0}\right)^{-\alpha} 
         \exp\left[-\left(\frac{E}{E_{\mathrm{cutoff}}}\right)^{\beta^\prime}\right],
\end{equation}
where $N(E)$ is the differential electron number density. The spectrum is defined using a reference energy $E_0 = 1\,\mathrm{GeV}$, normalization $A$, power-law index $\alpha$, and an exponential cutoff with characteristic energy $E_{\mathrm{cutoff}}$ and curvature parameter $\beta^{\prime}$. The magnetic field strength $B$ in astrophysical plasmas is commonly related to the thermal pressure $P$ through the plasma beta parameter:
\begin{equation}
\beta_p \equiv \frac{P}{P_B} = \frac{8\pi P}{B^2},
\end{equation}
where $P_B = B^2/8\pi$ is the magnetic pressure. This relation has been widely employed in studies of relativistic jets and AGN to estimate magnetic field strengths when direct measurements are unavailable \citep{1983MNRAS.205..449W,1995ApJ...449L..19G,2023ApJ...944..199S,2025MNRAS.539.3697D}. Assuming approximate pressure equilibrium between thermal and magnetic components, the field strength can be expressed as:
\begin{equation}
B = \sqrt{\frac{8\pi P}{\beta_p}}.
\end{equation}
The pressure equilibrium prescription was first introduced by~\citet{1983MNRAS.205..449W} for modeling synchrotron emission in extended radio sources. This approach has since been adapted for relativistic jet simulations \citep{2019MNRAS.483.3444P,2020Natur.582..356H,2025MNRAS.539.3697D}. The derived high parameter ($\beta_p \gg 1$) confirms the lobes are weakly magnetized, with thermal pressure dominating magnetic pressure.
\begin{figure*}
   \centering
    {\includegraphics[angle=0,width=0.49\textwidth]{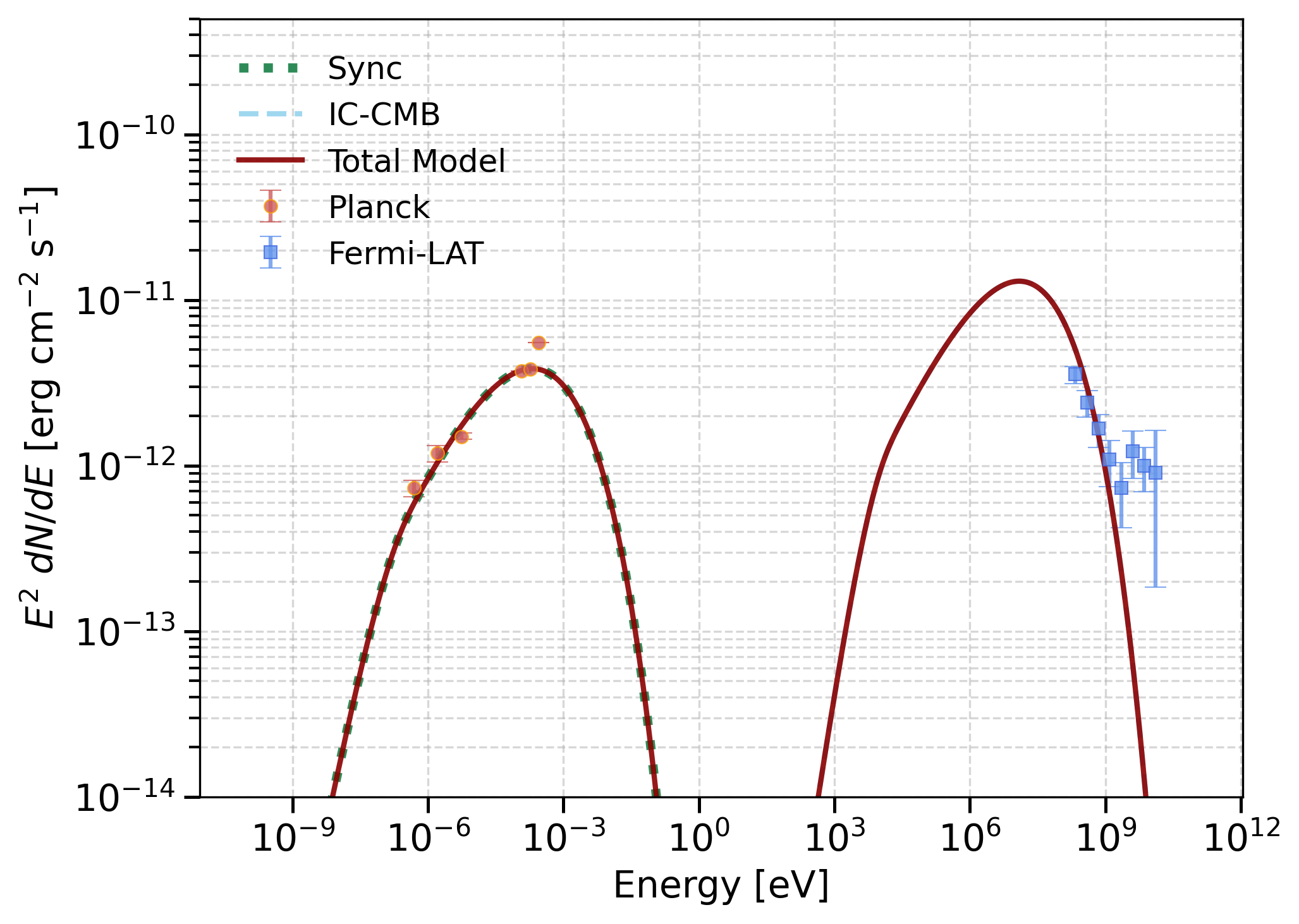}}
    {\includegraphics[angle=0,width=0.49\textwidth]{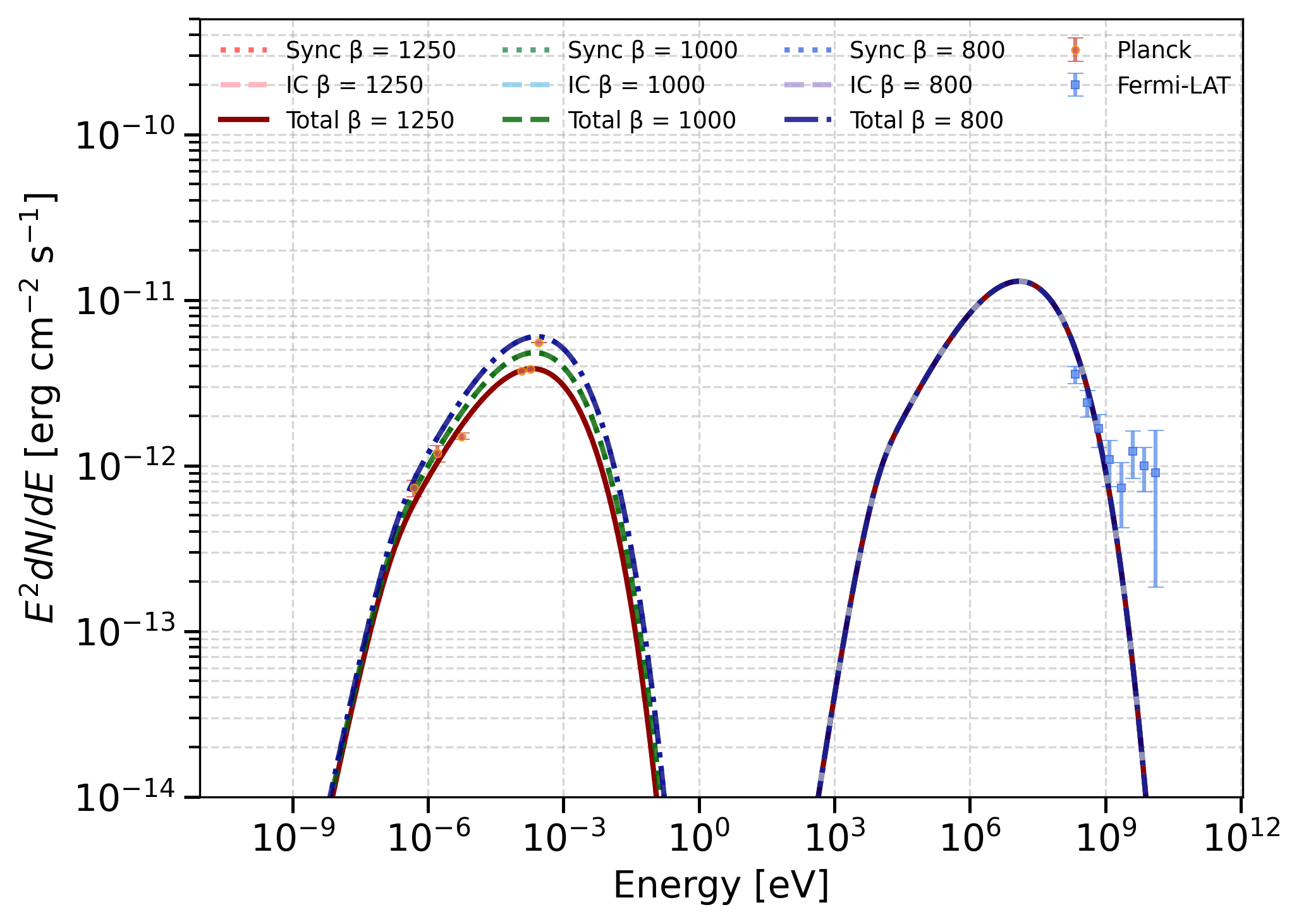}}  
  \caption{
   \textbf{Left:} Spectral energy distribution of Centaurus~A's inner lobes calculated using \textsc{naima} with parameters from~\citet{2016A&A...595A..29S}. The green dotted curve shows synchrotron emission from an exponential-cutoff power-law electron spectrum ($\alpha=1.79$, $E_{\rm cut}=21.2\,\mathrm{GeV}$) for $B=1.8\,\mu\mathrm{G}$ (corresponding to $\beta_p = 1250$). The green dashed curve displays the modeled synchrotron component, while the light-blue dashed curve corresponds to IC emission from CMB photon scattering. The combined synchrotron and IC components are shown as the maroon solid curve, representing the total predicted emission.
    \textbf{Right:} SED dependence on plasma-$\beta$ parameter showing solutions for $\beta_p=800$ -- $1250$, with $\beta_p=1250$ (maroon) providing the best match to observations.
  }
    \label{fig:SED}
\end{figure*}
Figure~\ref{fig:SED} presents the broadband emission properties of Centaurus~A's inner lobes, derived from our precessing jet simulation assuming $P=1.8\,\mathrm{Myr}$ and using the \textsc{naima} package~\citep{naima}. The derived magnetic field strength $B=1.8\,\mu\mathrm{G}$ corresponds to $\beta_p \approx 1250$, consistent with findings of low magnetic pressure in radio galaxy lobes, where plasma dominates over magnetic fields~\citep{2024A&A...683A.235R, 2014MNRAS.438.3310C, 2023A&A...679A.160O}. The jet with a $1.8\,\mathrm{Myr}$ precession period controls how energy is distributed, influencing the regions where particles are accelerated. The maximum electron energy used in our leptonic model, $E_{\rm cutoff} = 21.2\,\mathrm{GeV}$, is adopted from the observational SED modeling of the lobe by \citet{2016A&A...595A..29S}. This value represents a radiation-loss limit for electrons in the lobe's weak $\sim 1.8\,\mu\mathrm{G}$ magnetic field over the relevant timescale. The $\beta_p$ - dependence on Figure~\ref{fig:SED} - (right panel) demonstrates how moderate variations ($\beta_p=800$ -- $1250$) systematically affect the synchrotron emission, with $\beta_p=1250$ solution providing the best match to the observed broadband spectrum while remaining consistent with pressure estimates derived from X-ray data~\citep{2019MNRAS.485..872W}.

While the N3 region encompasses both lobes and jet knots, the compact parsec-scale nature of knots renders their contribution negligible in integrated analyses due to their locally intense $\sim 30 - 100\,\mu\mathrm{G}$ fields, which produce bright but spatially confined synchrotron emission that becomes spectrally diluted when averaged across the entire N3 volume~\citep{2008ApJ...673L.135W, 2010ApJ...708..675G, 2025MNRAS.539.3697D}. This explains why global SED fitting, which attributes gamma rays to IC scattering of lobe electrons, converges on weaker lobe field strengths~\citep{2013A&A...558A..19W}. The \textit{Fermi}-LAT results support this picture, showing spatially resolvable lobe emission contributing to the hard spectral component above $6.5\,\mathrm{GeV}$, distinct from the core's synchrotron self-Compton dominated emission~\citep{2019MNRAS.483.3444P}. Additionally, H.\,E.\,S.\,S. observations above $240\,\mathrm{GeV}$ exceed predictions of a single-zone exponential cutoff power-law model with $E_{\rm cutoff}=21.2\,\mathrm{GeV}$ by an order of magnitude~\citep{2018A&A...619A..71H, 2020Natur.582..356H, 2021MNRAS.505.1334W}, indicating a separate acceleration component is required for the TeV emission. Therefore, H.\,E.\,S.\,S. knots data are treated as evidence for a distinct high-energy electron population while focusing the one-zone lobe model on the bulk dynamics of Centaurus A's inner lobes.

\begin{figure}
   \centering
    \includegraphics[angle=0,width=0.4\textwidth]{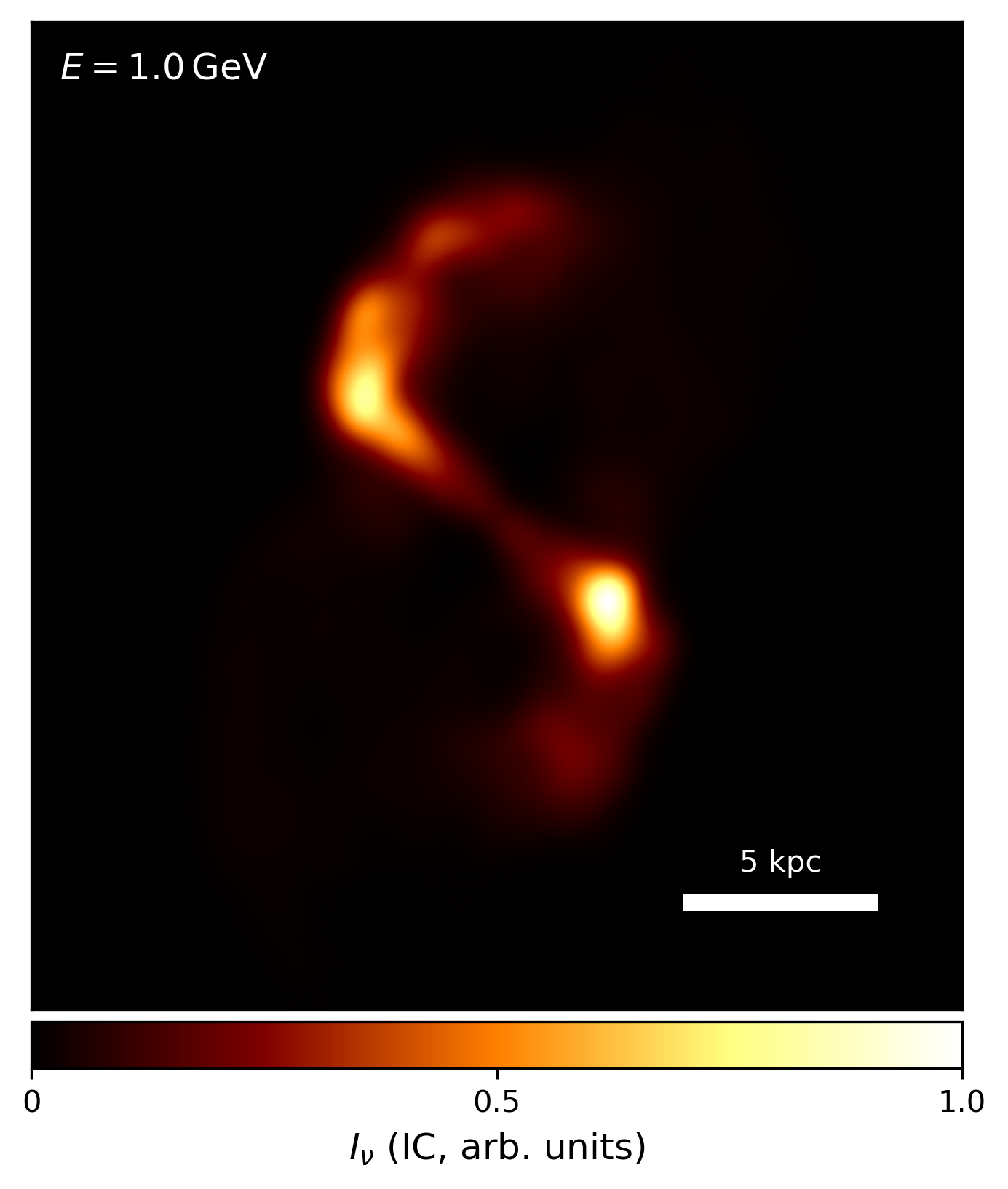}
  \caption{Simulated IC emission from Centaurus~A at $1.0\,\mathrm{GeV}$, viewed at $\theta_{\rm prec} = 50°$. The image shows the projected intensity integrated along the line of sight. The IC emissivity is calculated using a radiative model with an exponential cutoff power-law electron distribution interacting with CMB photons. Intensity in arbitrary units. }
    \label{fig:IC_image}
\end{figure}

In addition, the magnetic field strength of $B\sim1.8 \,\mu\mathrm{G}$ derived from our SED modeling is significantly lower than the value of $13\,\mu\mathrm{G}$ reported by \citet{1983ApJ...273..128B} and \citet{2021MNRAS.505.1334W} for Centaurus~A's inner lobes. This discrepancy suggests the lobe deviates strongly from minimum-energy conditions, with the plasma $\beta_p$ reaching $\sim 1250$. Such low $B$-field values align with X-ray observations of the lobe's thermal pressure (\(P_{\mathrm{th}} \sim 2\times10^{-10}\,\mathrm{dyn\,cm^{-2}}\)~\citep{2002ApJ...569...54K}), which dominates over the equipartition magnetic pressure (\(P_{B} \sim 10^{-12}\,\mathrm{dyn\,cm^{-2}}\)) by two orders of magnitude. The large $\beta_p$ value implies that the lobe is strongly dominated by particles, with thermal and cosmic-ray pressures overwhelming magnetic effects. This may result from precession-driven turbulence disrupting field amplification or from efficient particle injection during jet-ICM interactions. Similar sub-equipartition conditions in other FRI lobes~\citep{2005ApJ...626..733C,2007ApJ...669..893H} suggest that kinetic energy transport in low-power jets preferentially heats particles rather than amplifying fields, with precession-induced asymmetries potentially explaining the differing Faraday rotation signatures between northern and southern lobes~\citep{1983ApJ...273..128B}.

It is important to emphasize that the dominance of thermal pressure over magnetic pressure reported here applies specifically to the extended lobe plasma of the N3 region, and should not be conflated with conditions within the collimated jet spine. In the jet beam, radio polarization observations provide an upper limit on the thermal electron density, implying that nonthermal (relativistic electron) pressure typically exceeds thermal pressure, a well-established result for FRI jets~\citep{2002MNRAS.336..328L, 2005ApJ...626..733C}. Our RHD model does not contradict this, as the bulk pressure in our simulations represents an effective total pressure of the relativistic plasma, and cannot be decomposed into separate thermal and nonthermal contributions without explicit magnetic field and cosmic-ray transport equations. In the extended lobe, however, jet-ICM mixing and turbulent dissipation over Myr timescales are expected to increase the thermal and cosmic-ray pressure relative to the magnetic field, naturally producing the 
sub-equipartition conditions ($\beta_p \gg 1$) we derive. A self-consistent treatment of the pressure partition between thermal, 
nonthermal, and magnetic components across both the jet and the lobes will require future RMHD simulations with explicit cosmic-ray 
transport.

Figure~\ref{fig:IC_image} shows the simulated IC emission at $1.0\,\mathrm{GeV}$ for a jet oriented at an inclination angle $\theta_{\rm prec} = 50^\circ$. The emission is computed from a hydrodynamical simulation of the jet, scaled using a \textsc{naima}~\citep{naima} IC model with an exponential cutoff power-law electron distribution ($\alpha = 1.79$, $E_\mathrm{cut} = 21.2\,\mathrm{GeV}$~\citep{2016A&A...595A..29S}). The intensity image (arbitrary units) highlights regions where the Mach number exceeds $1.0$, with electron acceleration efficiency $f_e = 0.5$. A Gaussian smoothing ($\sigma = 10$ pixels) is applied to the integrated emissivity. Our simulated IC emission reveals a structured spatial distribution that aligns with key features observed in the inner lobes of Centaurus~A. The simulated emission shows brightened edges and offset intensity maxima, characteristic of shock-accelerated electrons interacting with photon fields. These features align with the double-peaked gamma-ray morphology observed by \textit{Fermi}-LAT (Figure~\ref{fig:cenA_inner_lobes}), where the $5.0\,\mathrm{kpc}$ spatial scale matches our simulated emission structure. The asymmetry between northern and southern lobes emerges naturally from our precessing jet model, consistent with known density variations in Centaurus~A's environment~\citep{1983ApJ...273..128B, 2002ApJ...569...54K}.

The simulated emission peaks in regions where the Mach number exceeds unity, reinforcing the role of shock acceleration at the jet-ICM interaction sites. These shocks are a natural consequence of jet precession, as seen in our velocity and pressure images (Figure~\ref{fig:3D_pressure_density}). While our hydrodynamic model reproduces the large-scale IC morphology, it does not fully capture finer details such as filamentary substructures or the very high-energy emission observed by H.\,E.\,S.\,S.~\citep{2020Natur.582..356H,2021MNRAS.505.1334W,2025MNRAS.539.3697D}. These limitations likely arise from the absence of magnetic fields and small-scale ICM inhomogeneities in our simulations. Future observations with the Cherenkov Telescope Array (CTAO) will test our predictions by resolving the IC emission above 10 TeV and distinguishing leptonic from hadronic contributions \citep{2019scta.book.....C}. To interpret these data and assess the role of precession in driving spectral and morphological asymmetries, RMHD simulations with anisotropic cosmic‐ray transport will be essential.

\section{Summary and Conclusions}
\label{sec:conclusion}

In this work, we performed three-dimensional relativistic hydrodynamic simulations of precessing jet in Centaurus~A, using observationally motivated parameters. Our key findings are as follows: Jet precession reproduces observed morphology: A precession period of $1.8\,\mathrm{Myr}$, with a precession angle of \( 50^{\circ} \) yields synthetic radio emission images that closely match the observed S-shaped structure and lobe asymmetries of Centaurus~A’s inner radio lobes (Figure~\ref{fig:compare_sim_obs}).

The simulations capture broad, laterally extended lobes, curved jet trajectories, and limb-brightened emission, all of which are consistent with VLA radio observations~\citep{1992ApJ...395..444C}. The 3D simulations show that precession drives cocoon asymmetries, internal shocks, and turbulent mixing, which modify the velocity and thermodynamic structure of the jet and its environment. These effects align with observed X-ray shells and radio knots~\citep{2009MNRAS.395.1999C}. The Mach number and bulk thermal gas temperature distributions, in particular, indicate the presence of internal shocks and localized pressure gradients, features that may account for the observed asymmetries and filamentary substructure in the inner lobes of Centaurus~A~\citep{2008ApJ...673L.135W,2025MNRAS.539.3697D}. The results demonstrate that precession not only reshapes the large-scale morphology of the jet but also strongly modulates the thermodynamic properties of the surrounding medium.

The quantitative analysis of curvature, point symmetry, and lobe misalignment demonstrates that the jet evolution is consistent with precessional motion. At the end of one full precession cycle ($1.8\,\mathrm{Myr}$), the simulations yield \( C \geq 1000 \), indicating significant curvature; \( S \) fluctuates beyond \( \pm 0.05 \), reflecting S-shaped symmetry between the lobes; and \( E \) reaches values \( \geq 0.2 \), confirming substantial off-axis displacement. These markers reinforce the hypothesis of a slowly reorienting jet axis over Myr timescales. 

The derived magnetic field strength of $\sim 1.8 \,\mu\mathrm{G}$ ~\citep{2016A&A...595A..29S} corresponds to a plasma beta parameter $\beta_p \approx 1250$, indicating that the lobes are weakly magnetized and dominated by thermal and cosmic-ray pressures, which aligns with observations of other FRI radio galaxies where particle pressure often overwhelms magnetic pressure. The IC emission at 1 GeV, simulated for a precessing angle $\theta_{\rm prec} = 50^\circ$, shows a structured spatial distribution that matches the double-peaked gamma-ray morphology observed by \textit{Fermi}-LAT~\citep{2019MNRAS.483.3444P}, supporting the idea that precession shapes the large-scale emission features. The observed asymmetry in gamma-ray flux between the northern and southern lobes ($F_{\gamma,\text{SW}} \approx 2 \times F_{\gamma,\text{NE}}$) is naturally explained by environmental differences, such as denser gas surrounding the southern lobe, as suggested by X-ray observations~\citep{2003ApJ...592..129K}, which is consistent with Faraday depolarization measurements indicating that the southern lobe lies behind a clumpy intracluster medium.

Overall, our results suggest that the complex inner lobe morphology of Centaurus~A is shaped by gradual, long-term jet precession, likely driven by either binary black hole dynamics or instabilities in the accretion disk. While the model reproduces the large-scale S-shaped morphology and asymmetries observed in the inner lobes, it neglects several important physical processes, including magnetic field dynamics, radiative cooling, and turbulent instabilities. The simulations also assume a uniform hydrodynamic treatment of the jet–ICM interaction, which limits their ability to resolve small-scale features such as filamentary substructures, polarization patterns, and localized shock acceleration zones, particularly evident in the southern lobe. These limitations underscore the need for more comprehensive models. Future studies incorporating magnetohydrodynamic effects, radiative processes, and jet-ICM will yield deeper insights into the small-scale structure and energy dynamics of Centaurus~A.

\section*{Acknowledgements}

The authors are grateful to Geoffrey Bicknell for insightful comments and suggestions that significantly improved the clarity and presentation of this paper.
The authors acknowledge S. Boula and JS. Wang for helpful discussions. We gratefully thank Dmitry Prokhorov for generously sharing the data used to reproduce Figure \ref{fig:cenA_inner_lobes} in this paper. R.C.A. research is supported by CAPES/Alexander von Humboldt Program (88881.800216/2022-01). The simulations presented here were performed on the HPC system Raven at the Max Planck Computing and Data Facility. R.C.A. gratefully acknowledges the Max Planck Institute for Nuclear Physics for their warm hospitality and support during her visit, which provided a conducive environment for fruitful discussions and collaborations. 

\section*{Data Availability}

The data supporting this study are available from the corresponding author upon reasonable request.



\bibliographystyle{mnras}
\bibliography{example} 

\bsp	
\label{lastpage}
\end{document}